\documentclass{article}
\usepackage{spconf,amsmath,graphicx}
\usepackage{tabularx}
\usepackage{makecell}
\usepackage{multirow}
\usepackage{color}
\usepackage{amssymb}
\usepackage{enumitem}
\setlist{nosep, leftmargin=14pt}

\usepackage{mwe} % to get dummy images

\title{Spatially Exclusive Pasting: A General Data Augmentation \\ for the Polyp Segmentation}
\name{Lei Zhou}
\address{State Key Laboratory for Novel Software Technology, Nanjing University, China}
\begin{document}
%\ninept
%
\maketitle

\begin{abstract}
    
    Automated polyp segmentation technology plays an important role in diagnosing intestinal diseases, such as tumors and precancerous lesions. Previous works have typically trained convolution-based U-Net or Transformer-based neural network architectures with labeled data. However, the available public polyp segmentation datasets are too small to train the network sufficiently, suppressing each network's potential performance. To alleviate this issue, we propose a universal data augmentation technology to synthesize more data from the existing datasets. Specifically, we paste the polyp area into the same image’s background in a spatial-exclusive manner to obtain a combinatorial number of new images. Extensive experiments on various networks and datasets show that the proposed method enhances the data efficiency and achieves consistent improvements over baselines. Finally, we hit a new state of the art in this task. We will release the code soon.

% {\bf\emph{ Index Terms-\ Polyp Segmentation; Data Augmentation; Convolution; Transformer; }\rm}
\end{abstract}

\begin{keywords}
Polyp Segmentation; Data Augmentation; Convolution; Transformer;
\end{keywords}

\begin{figure}[t]
\begin{minipage}[b]{0.45\linewidth}
  \centering
  \centerline{\includegraphics[width=4.0cm]{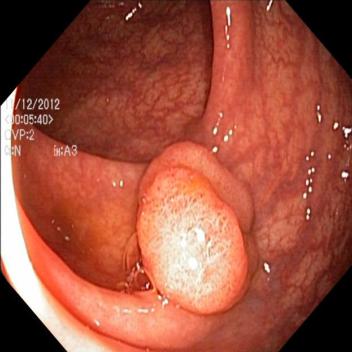}}
%  \vspace{2.0cm}
  \centerline{(a) Original Image}\medskip
\end{minipage}
\begin{minipage}[b]{0.45\linewidth}
  \centering
  \centerline{\includegraphics[width=4.0cm]{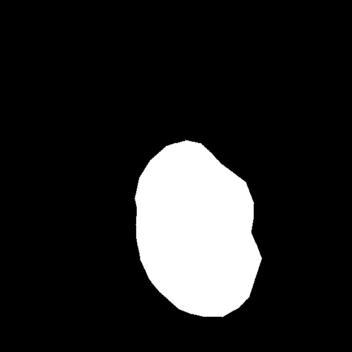}}
%  \vspace{1.5cm}
  \centerline{(b) GT}\medskip
\end{minipage}
\begin{minipage}[b]{0.45\linewidth}
  \centering
  \centerline{\includegraphics[width=4.0cm]{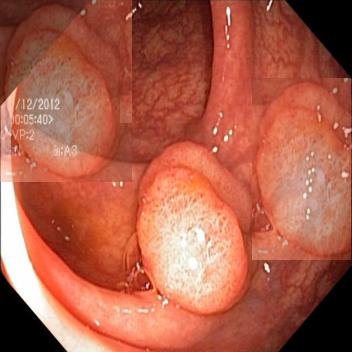}}
  \label{fig:augimg}
%  \vspace{2.0cm}
  \centerline{(c) Augmented Image}\medskip
\end{minipage}
\begin{minipage}[b]{0.62\linewidth}
  \centering
  \centerline{\includegraphics[width=4.0cm]{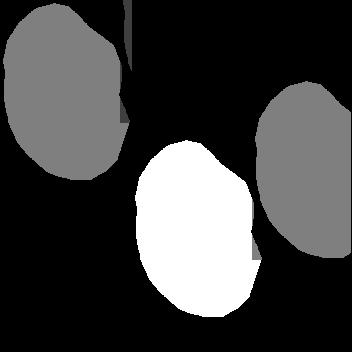}}
%  \vspace{1.5cm}
  \centerline{(d) Augmented GT}\medskip
\end{minipage}
\hfill
\vspace{-0.5cm}
\caption{An example of the augmentation results. The shadows in the augmented GT indicate the locations to paste the cropped polyp.}
\label{fig:res}
\end{figure}

\vspace{-0.2cm}
\section{Introduction}
\vspace{-0.2cm}
Polyp recognition holds an essential place in the process of colonoscopy. Automated polyp segmentation with deep learning helps doctors to reduce missed cases. With the aid of pixel-wise labeled data, previous works devoted to designing more and more complex network architectures, like the convolution-based U-Net~\cite{resunet, pranet,acsnet,msrf,dcrnet} and Transformer-based network~\cite{ssformer, FCBFormer,colonformer,transfuse}, to learn better representations for the input images. 

\begin{figure*}[t]
    \centerline{\includegraphics[width=15cm]{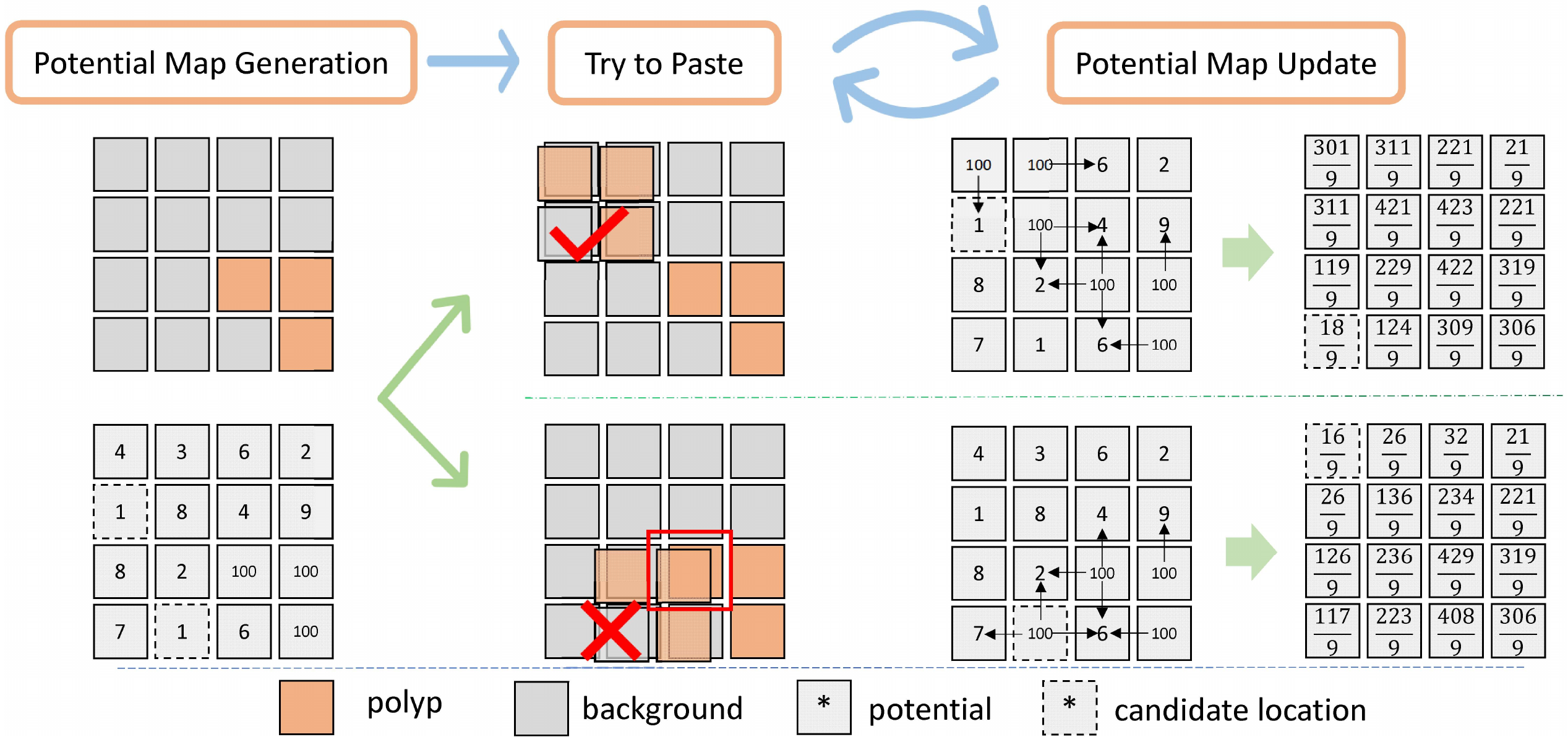}}
\caption{The process of Spatially Exclusive Pasting.}
\label{fig:furtherpaster}
\end{figure*}

These networks are often data-hungry. However, there only exist small datasets for this task since labeling polyp area relies heavily on experienced experts. Thus, proper data augmentation technology impacts the results significantly. Some geometric transformations, like \emph{Random Flip}, \emph{Random Rotate}, and \emph{Random Zoom}, are often applied as basic components. Yet, they are designed for general purposes and not specifical for polyp segmentation. Mixup~\cite{mixup} is a widely used augmentation that generates a new image by linearly combining two input images and supervises the new image with the linearly combined labels. Some of its variants like the CutOut~\cite{cutout} and CutMix~\cite{cutmix} have also achieved promising results. Unfortunately, they are designed for classification in nature and their procedures are object-agnostic, which violates the object awareness in the segmentation task. Copy-Paste~\cite{copypaste} improves the instance segmentation task~\cite{maskrcnn} by copying the instance in one image and randomly pasting it to the other image to obtain augmented data. Although this method meets most of our requirements, there still exist two critical differences between polyp segmentation and instance segmentation. First, in polyp segmentation, the foreground and background look very similar and identifying the polyp area relies on carefully modeling the contextual relations to the current background, while the objects in instance segmentation are more self-identifiable. This causes two problems: 1. only pasting the foreground area is insufficient since it loses context information for polyp segmentation; 2. it's hard for the current polyp to build effective contextual relations with other images' backgrounds. Second, unlike the diverse foregrounds in instance segmentation, polyp segmentation only owns one category of the foreground. Thus not only is it impossible for the overlapped foreground area caused by the random pasting to provide the desired convex combination of different categories, but it also may damage the original appearance of the polyp area, which confuses the learning process of foreground representation.

To handle these issues, we propose a novel and polyp-segmentation-specific data augmentation, so-called Spatially Exclusive Pasting, which differs from the Copy-Paste~\cite{copypaste} in three aspects: 1. we copy and paste 
the foreground along with the surrounding; 2. the copying and pasting are carried out in the same image; 3. rather than randomly choosing pasting location, we determine the target area using a potential map guided strategy. This strategy forbids the foregrounds to overlap and makes the pasted foreground keep away from each other. Extensive experiments on multiple backbones, including the previous state-of-the-art model, show solid improvements over the baselines. This indicates that the proposed method is a universal augmentation for polyp segmentation.

% To address these issues, we construct a unified framework in which different methods are trained using the exact same settings, as our software platform to conduct all of our following experiments. Then, to alleviate the shortage of data, we propose a novel and universal data augmentation strategy, so-called Farthest Patch Paste, to exploit the information within existing datasets. Specifically, for each sample, we crop the foreground patch and randomly paste them into spatial-exclusive area to generate new training sample, as depicted in Fig~\ref{fig:paste}. This effectively lifts the ratio of the polyp areas and enrich the combinations of the polyp and the healthy part. Based on the above, we conduct extensive experiments. The results shows that the proposed data augmentation improve both categories network well 

\vspace{-0.3cm}
\section{Methodology}
\vspace{-0.2cm}
In this section, we will detail the proposed Spatially Exclusive Pasting. The whole process is depicted in Fig~\ref{fig:furtherpaster}.

% \subsection{Farthest Patch Paste}

For a given image $I\in \mathbb{R^{H\times W\times 3}}$ and the corresponding label map $L\in \mathbb{R^{H\times W}}$, we denote the cropped foreground patch as $I_f$ and its label map as $L_f$. Suppose that 

\begin{align}
    h_s, w_s=&\mathop{\arg\min}(L==1),\\ 
    h_e, w_e=&\mathop{\arg\max}(L==1),
\end{align}

\noindent
where 1 in the label map indicates the polyp, we have

\begin{align}
    I_f = I[h_s:h_e, w_s:w_e],\\
    L_f = L[h_s:h_e, w_s:w_e].
\end{align}

The basic idea of the proposed method for generating new images is very simple: copy the polyp area and paste it into other areas to obtain a combinatorial number of new images. The key concern is where to paste. A trivial solution is to pick locations randomly. However, it may cause conflict in the original foreground area. A cumbersome solution is to calculate the farthest location to the polyp area to avoid overlap. But this means would lack randomness and introduces incorrect polyp distribution bias. Thus, we design a potential map-based process to raise candidate locations. The whole augmentation process consists of three modules: a potential map generation module to initialize each coordinate with a potential, a pasting module and an update module to update the potential of each coordinate.

\begin{table*}[tbp]
\begin{center}
    \caption{Results on Kvasir-SEG. 'w/ SEP' denotes to train with the Spatially Exclusive Pasting strategy.}\label{tab:kvasir}
\begin{tabular}{c|c|cc|c|c|cc}
\hline\hline
Seed & Models & mDice$(\%)\uparrow$ & mIoU$(\%)\uparrow$ & Seed& Models &  mDice$(\%)\uparrow$ & mIoU$(\%)\uparrow$ \\ \hline
\multirow{8}{*}{0}  & 
ResUNet~\cite{resunet} &92.91&88.21  
&\multirow{8}{*}{2} 
&ResUNet~\cite{resunet}&92.04&87.33 \\
                    & w/ SEP &\textcolor[rgb]{0,0,1}{(+0.59)}93.5&\textcolor[rgb]{0,0,1}{(+0.71)}88.92 && w/ SEP &\textcolor[rgb]{0,0,1}{(+0.63)}92.67&\textcolor[rgb]{0,0,1}{(+0.88)}88.21\\ 
                     \cline{2-4} \cline{6-8}
                    & PraNet~\cite{pranet} &93.02&88.68 && PraNet~\cite{pranet} &92.74&88.08  \\ 
                    & w/ SEP &\textcolor[rgb]{0,0,1}{(+0.74)}93.76&\textcolor[rgb]{0,0,1}{(+0.71)}89.39 && w/ SEP &\textcolor[rgb]{0,0,1}{(+0.26)}93.00&\textcolor[rgb]{0,0,1}{(+0.41)}88.49 \\ 
                    \cline{2-4} \cline{6-8}
                    & SSFormer-S~\cite{ssformer}&94.18&90.13 & & SSFormer-S~\cite{ssformer}&93.72&89.6 \\
                    & w/ SEP &\textcolor[rgb]{0,0,1}{(+0.31)}94.49&\textcolor[rgb]{0,0,1}{(+0.42)}90.55 && w/ SEP &\textcolor[rgb]{0,0,1}{(+0.75)}94.47&\textcolor[rgb]{0,0,1}{(+0.77)}90.37\\ 
                    \cline{2-4} \cline{6-8}
                    % & SSFormer-L~\cite{ssformer} &94.46&90.73 & & SSFormer-L~\cite{ssformer} &93.9&89.92\\ 
                    % & w/ SEP &94.23&90.19 & & w/ SEP &93.71&89.5\\ 
                    \cline{2-4} \cline{6-8}
                    & FCBFormer~\cite{FCBFormer}&93.93&89.56 & & FCBFormer~\cite{FCBFormer}&94.32&90.17\\
                    & w/ SEP &\textcolor[rgb]{0,0,1}{(+0.86)}94.79&\textcolor[rgb]{0,0,1}{(+1.16)}90.72 & & w/ SEP &\textcolor[rgb]{0,0,1}{(+0.25)}94.57&\textcolor[rgb]{0,0,1}{(+0.68)}90.85\\
                    \hline\hline
\multirow{8}{*}{1}  & ResUNet~\cite{resunet} &90.49&85.09  &\multirow{8}{*}{fix} &ResUNet~\cite{resunet} &92.72&88.41 \\
                    & w/ SEP &\textcolor[rgb]{0,0,1}{(+0.05)}90.54&\textcolor[rgb]{0,0,1}{(+0.14)}85.23 && w/ SEP &\textcolor[rgb]{0,0,1}{(+0.4)}93.12&\textcolor[rgb]{0,0,1}{(+0.48)}88.89\\ 
                    \cline{2-4} \cline{6-8}
                    & PraNet~\cite{pranet} &91.45&86.34 && PraNet~\cite{pranet} &92.31&87.73  \\ 
                    & w/ SEP &\textcolor[rgb]{0,0,1}{(+0.23)}91.68&\textcolor[rgb]{0,0,1}{(+0.36)}86.7 && w/ SEP &\textcolor[rgb]{0,0,1}{(+0.82)}93.13&\textcolor[rgb]{0,0,1}{(+0.73)}88.46\\ 
                    \cline{2-4} \cline{6-8}
                    & SSFormer-S~\cite{ssformer}&92.31&87.23 & & SSFormer-S~\cite{ssformer}&93.19&89.05 \\
                    & w/ SEP &\textcolor[rgb]{0,0,1}{(+0.45)}92.76&\textcolor[rgb]{0,0,1}{(+0.6)}86.51 && w/ SEP &\textcolor[rgb]{0,0,1}{(+0.65)}93.84&\textcolor[rgb]{0,0,1}{(+0.72)}89.77\\ 
                     \cline{2-4} \cline{6-8}
                    % & SSFormer-L~\cite{ssformer} &92.40&87.06 & & SSFormer-L~\cite{ssformer} &93.57&89.43\\ 
                    % & w/ SEP &\textcolor[rgb]{0,0,1}{(+0.51)}92.91&\textcolor[rgb]{0,0,1}{(+0.91)}87.97 & & w/ SEP &\textcolor[rgb]{0,0,1}{(+0.42)}93.99&\textcolor[rgb]{0,0,1}{(+0.85)}90.28\\ 
                     \cline{2-4} \cline{6-8}
                    & FCBFormer~\cite{FCBFormer}&91.93&86.73 & & FCBFormer~\cite{FCBFormer}&93.91&89.84\\
                    & w/ SEP &\textcolor[rgb]{0,0,1}{(+0.53)}92.46&\textcolor[rgb]{0,0,1}{(+0.79)}87.52 & & w/ SEP &\textcolor[rgb]{0,0,1}{(+0.2)}94.11&\textcolor[rgb]{0,0,1}{(+0.18)}90.02\\ 
                    \hline\hline
\end{tabular}
\end{center}
\end{table*}

First, the generation module randomly initializes a map $\mathbf{M}\in\mathbb{R}^{H\times W}$ of the same size as the input image. The coordinate $(p_x, p_y)$ with the lowest value will be raised to be the candidate location for pasting. We manually set the locations corresponding to the foreground to a large value $T$, and thus the foreground area is kept from being raised.

Then, we tentatively paste the cropped foreground to the candidate area $c=[p_x:p_x+h_e-h_s, p_y:p_y+w_e-w_s]$. If there is an overlap of polyp between the cropped patch and the pasting location, we revoke the pasting and lift the potential value of this pasting location to the large value $T$.  If no overlaps, we replace the candidate area with the following convex combination:

\begin{align}
    I[c] &= \alpha \times I[c]+(1-\alpha)\times I_f, \\
    L[c] &= \alpha\times L[c]+(1-\alpha)\times L_f,
\end{align}

\noindent
where $\alpha=0.7$. The potential map is updated by 

\begin{equation}
    M[c] = T.
\end{equation}

After the tentative pasting, some potential values are changed to a large value $T$, which means they will never be selected as candidates. Based on a reasonable assumption that the nearby coordinates share a similar likelihood of avoiding overlap, we spread the potential value of each coordinate to its neighbors with the following convolution operation:

\begin{equation}
    M = conv(M, w),
\end{equation}
\noindent
where $w$ is a mean filter, e.g.,

\begin{equation}       
w=\left[
  \begin{array}{ccc} 
    1/9 & 1/9 &1/9\\
    1/9 & 1/9 &1/9\\
    1/9 & 1/9 &1/9
  \end{array}
\right]
\end{equation}

The pasting module and the update module of the potential map are implemented iteratively 10 times. Thus the foreground area may appear in multiple locations, and an example is depicted in Fig~\ref{fig:res}.

\vspace{-0.2cm}
\section{Experiments}
\vspace{-0.2cm}
To verify the effectiveness of the proposed methods, we conduct extensive experiments on the different baseline models in this section. To further explore its characteristics, we also provide ablation studies.

\subsection{Implementation Details}
All of the experiments are executed on a server with one 2080Ti GPU. The whole project is implemented using PyTorch. We train every model for 150 epochs with an initial learning rate of 1e-4. During training, the learning rate is scheduled by the CosineAnnealingLR~\cite{coslr}, with the first 5 epochs to warm up. We adopt gradient clipping to restrict the norm of the gradient to be lower than 20, which stabilizes the training process greatly. And the network parameters are updated with the default Adam optimizer~\cite{adam}. We still utilize some geometrical data augmentation in our experiments, including the random horizontal and vertical flip, random rotation and random zoom. We will show that our proposed augmentation is additive to these basic augmentations.

\vspace{-0.3cm}

\subsection{Datasets}

Following the common practice, we conduct experiments on two public polyp datasets, the Kvasir-SEG~\cite{kvasir} and EndoScene~\cite{endoscene}. They have 1000 and 912 labeled images, respectively. EndoScene~\cite{endoscene} has been partitioned into train/validation/test subset officially. As to the Kvasir-SEG, prior works~\cite{ssformer,FCBFormer} randomly divide it into train/validation/test subset using the rate of 8/1/1. However, according to our experiments, the minor differences in data partition cause great gaps in the evaluation results. Therefore we specify three random seeds [0,1,2] to split the dataset with sklearn. To further ensure reproducibility in different platforms, we also adopt a fixed style that sorts the samples according to file names and extracts the first 80\% samples as the training sets, the last 10\% as the validation set, and the remainings as the testing set. So, for each model in each dataset, we will provide 4 results, e.g. results for partition [0, 1, 2, fix].

\vspace{-0.3cm}

\subsection{Evaluations}

We report two convincing metrics: the mDice and mIoU. The results of every baseline are obtained through reproductions on our machine. Thanks to the stable experimental settings, all the baselines achieve better results than their reported ones with a certain random seed. With the aid of the proposed augmentation, they all achieve consistent improvements over baselines under all the random seed settings.

\vspace{-0.3cm}
\subsubsection{Baselines}
\vspace{-0.2cm}
Our method is tested on various representative baselines. They cover the key evolutionary stages in this task. 
\begin{itemize}
    \item ResUNet~\cite{resunet}, constructed with the convolution operations.
    \item PraNet~\cite{pranet}, constructed with convolutions and augmented by the attention module.
    \item SSFormer-S~\cite{ssformer}: constructed with the Transformer blocks.
    % \item SSFormer-L~\cite{ssformer}: Transformer-based network with more parameters.
    \item FCBFormer~\cite{FCBFormer}: the combination of convolution and Transformer branch.
\end{itemize}

\vspace{-0.4cm}
\subsubsection{Results on Kvasir-SEG}
\begin{figure}[t]
    \centerline{\includegraphics[width=8cm]{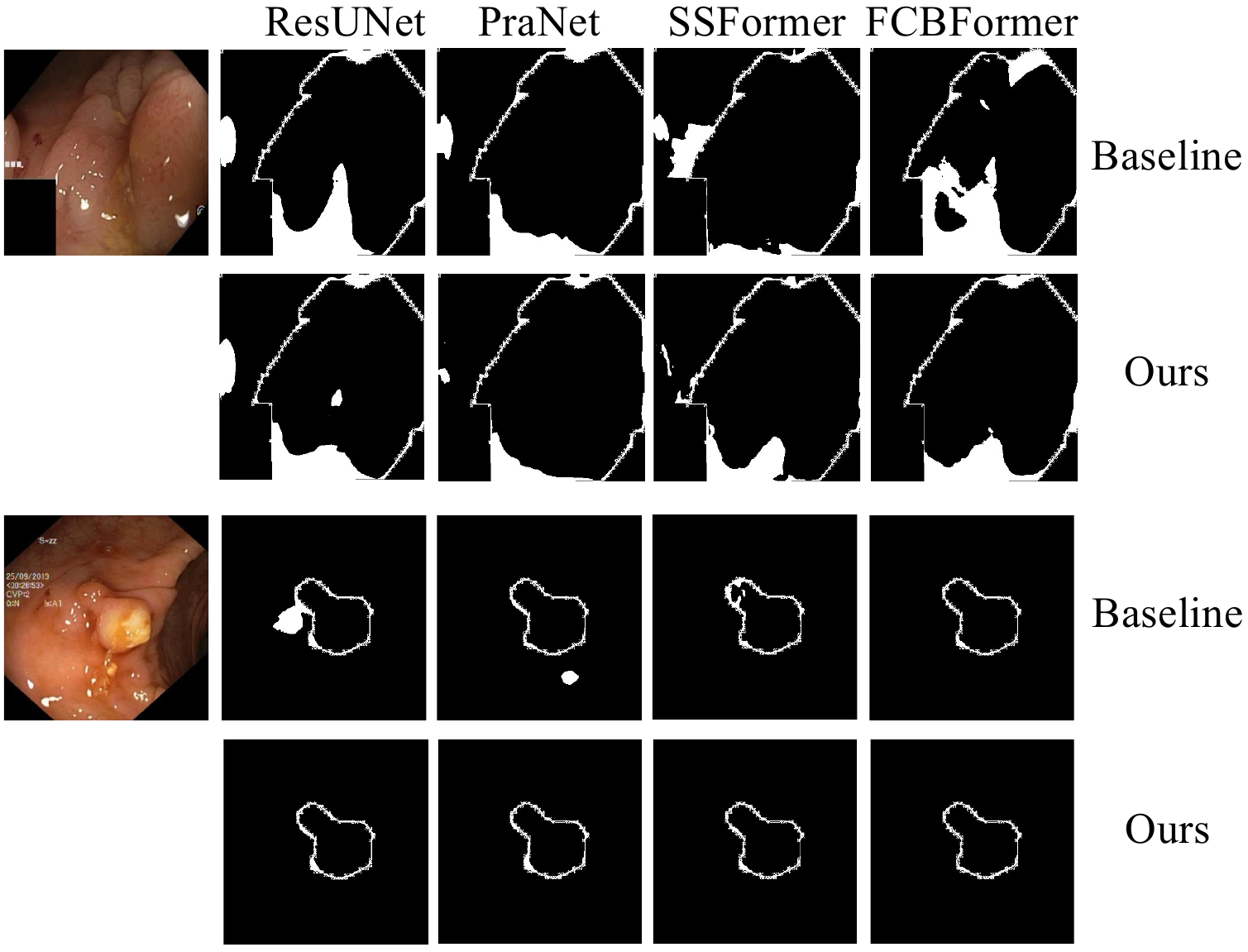}}
\vspace{-0.3cm}
\caption{Error map visualization. The white area is obtained by judging whether the prediction equals the ground truth. Our method improves the boundary areas.}
\label{fig:errormap}
\end{figure}

Table~\ref{tab:kvasir} shows the results of Kvasir-SEG in distinct backbones using assorted random seeds. Since the variances among different seeds are too big to ignore, we choose not to merge them into a "mean-variance" style. Our method boosts all the baselines under every random seed. This verifies that our method effectively improves the data efficiency and is a general augmentation in the polyp segmentation task. Meanwhile, we find that, with proper training, the gap between the newest and the traditional models is rather limited. Qualitative results are illustrated in Fig~\ref{fig:errormap}.

\vspace{-0.4cm}

\subsubsection{Results on EndoScene}
Since EndoScene~\cite{endoscene} provides the official partition, we directly adopt it. And we specify a fixed random seed,e.g. 0, during the data augmentation. Table~\ref{tab:cvcdb} shows that our methods balance the performance of the different networks. The PraNet~\cite{pranet} seems not to be benefited from the augmentation. We think it is because the training set of EndoScene~\cite{endoscene} merely contains 547 images, which makes it hard to train its attention module effectively. How to handle the attention module with a small dataset remains to be explored.

\begin{table}[tbp]
\begin{center}
    \caption{Results on EndoScene.}\label{tab:cvcdb}
\begin{tabular}{c|cc}
\hline
Methods &  mDice$(\%)\uparrow$ & mIoU$(\%)\uparrow$ \\ \hline

ResUNet~\cite{resunet} &86.02&79.34\\
w/SEP &\textcolor[rgb]{0,0,1}{(+1.2)}87.22&\textcolor[rgb]{0,0,1}{(+1.3)}80.64\\\hline
PraNet~\cite{pranet} &86.09&79.47\\
w/SEP &\textcolor[rgb]{0,0,1}{(+0.09)}86.18&\textcolor[rgb]{1,0,0}{(-0.02)}79.45\\\hline

SSFormer-S~\cite{ssformer} &86.95&80.67\\
w/SEP &\textcolor[rgb]{0,0,1}{(+0.25)}87.2&\textcolor[rgb]{0,0,1}{(+0.18)}80.85\\\hline

\end{tabular}
\end{center}
\end{table}

\begin{table}[tbp]
\begin{center}
    \caption{Ablations on the pasting manners.}\label{tab:potentialmask}
\begin{tabular}{c|cc}
\hline
Configs &  mDice$(\%)\uparrow$ & mIoU$(\%)\uparrow$ \\ \hline
No Paste &93.19&89.05\\
Random Paste &84.31&78.15\\
Non-overlap Paste &93.28&88.86 \\
Foreground-Only Paste  &93.39&89.12\\
Cross-Frame Paste &93.37&89.03 \\
Spatially Exclusive Paste &\textbf{93.84}&\textbf{89.77}\\ \hline

\end{tabular}
\end{center}
\end{table}

\vspace{-0.2cm}
\subsection{Ablation Study}
%\vspace{-0.2cm}
To further explore the characteristics of the method, we ablate the pasting process carefully. First, we study the pasting location. We replace the potential map-based pasting with random pasting or random non-overlap pasting. Random pasting means randomly picking the pasting location, and the random non-overlap pasting adds a restriction to forbid pasting in the foreground area. We try to paste the foreground only as suggested in previous work~\cite{copypaste}. And we also attempt to conduct inter-frame pasting. Then we investigate the effect of these pasting styles. Results in Table~\ref{tab:potentialmask} show that our potential map-guided strategy outperforms all the other variants. We attribute this to reducing spatial conflict among the polyp areas and keeping the original context.

\vspace{-0.2cm}
\section{Compliance with Ethical Standards}
\vspace{-0.2cm}
This is a numerical simulation study for which no ethical approval was required.

\vspace{-0.3cm}

\section{Conclusion}
\vspace{-0.2cm}
In this work, we propose a novel data augmentation strategy to alleviate the shortage of data for the polyp segmentation task. Extensive experiments verify that the proposed method consistently improves the performance of different networks. In the future, we will work on extending this method to more tasks and explore how to learn from various data sources.

\bibliographystyle{IEEEbib}
\bibliography{strings,refs}

\end{document}